\documentclass[preprint,1p,times,authoryear,twocolumn]{elsarticle}
\usepackage{amsmath,amssymb,booktabs,graphicx,siunitx,longtable}
\usepackage{microtype}
\usepackage{xurl}
\usepackage{subcaption}
\usepackage{float}
\usepackage{lineno}
\usepackage{setspace}
\usepackage{xcolor}
\usepackage{placeins}
\usepackage[colorlinks=true,allcolors=blue]{hyperref}
\newcommand{\dd}{\mathrm{d}}
\newcommand{\Rey}{\mathrm{Re}}

\journal{Coastal Engineering}

\begin{document}

\begin{frontmatter}
\title{Dopamine: A Cartesian grid large-eddy-simulation framework for coastal ocean modelling of free-surface waves}
\author[1]{Akshay Patil}
\affiliation[1]{organisation={3D Geoinformation Research Group, Faculty of Architecture and Built Environment, Delft University of Technology},
addressline={Julianalaan 134},
city={Delft},
postcode={2628BL},
country={The Netherlands}
}
\begin{highlights}
\item A low numerical dissipation solver simulates free-surface waves and breaking on Cartesian grids.
\item Floating and sinking particles are tracked through the air--water interface.
\item The solver reaches 77\% strong-scaling efficiency on 512 CPU cores.
\end{highlights}

\begin{abstract}
We present a Cartesian staggered-grid large-eddy simulation framework for modelling coastal ocean waves using the \textit{Dopamine} solver.
  The framework combines a geometric volume-of-fluid method for simulating free-surface wave evolution with one-way-coupled Lagrangian point-particle tracking for heavy and buoyant particles.
  A ghost-cell immersed-boundary method represents stationary solid boundaries, including the seabed and coastal obstacles, and their interaction with free-surface waves.
  The framework includes the Vreman and anisotropic minimum-dissipation (AMD) sub-grid-scale models for simulating near-wall turbulence.
  Comparisons with analytical solutions and established experimental and numerical benchmarks illustrate the framework's ability to reproduce wave dynamics and particle motion.
  The results provide promising insights into the framework's capabilities for coastal ocean modelling applications focused on investigating small-scale turbulent physics using a low-numerical-dissipation approach.
  The solver runs on multi-CPU architectures, with two-fluid strong- and weak-scaling efficiencies at 512 ranks, relative to 64 ranks, of up to 77\% and 73\%, respectively, and a single-GPGPU build has been verified against the CPU solver.
\end{abstract}
\end{frontmatter}

%\linenumbers
\onehalfspacing

\section{Introduction}
\label{sec:introduction}
Coastal ocean surface flows involve interacting physical processes, including wave breaking~\citep{ting1994}, turbulent mixing~\citep{drazen2009}, and sediment transport~\citep{nielsen1992}.
Wave-induced transport also affects floating marine litter, whose drift depends on its response to the wave motion~\citep{calvert2021}. 
Surface waves produce a net transport that cannot be inferred from the local mean velocity alone.
Although the leading-order motion beneath a periodic wave is oscillatory, fluid parcels undergo a mean displacement over successive cycles.
The difference between the Lagrangian and Eulerian mean velocities is the Stokes drift, which provides a reference for understanding wave-induced transport~\citep{vandenbremer2017}.
The measurements and model of~\citet{calvert2021} show how this mechanism can increase the drift of floating marine litter relative to that of fluid tracers.
Predicting particle transport therefore requires a description of both the wave kinematics and the particle response to those kinematics.
Additionally, coastal ocean flows exhibit multi-scale turbulent features as they transition from deep-water, largely inviscid flow regimes to strong turbulence induced by wave breaking and bottom-wall interactions.

Consequently, modelling the coastal ocean requires a numerical framework that can accommodate a moving multiphase interface (air--water), particle motion, and a fast and accurate solution algorithm for the underlying governing equations.
The multiphase interface can be represented using the volume-of-fluid (VOF) method, which describes the interface through the local volume fraction of each phase~\citep{hirt1981}.
The air--water interface is then reconstructed geometrically from the volume fraction field occupied by the liquid phase~\citep{rider1998,weymouth2010}, making this approach suitable for studying flows that extend from regular waves to breaking.

For such a multiphase system, accurate interface transport must be accompanied by a consistent momentum balance across the large density contrast, and thus a non-constant-coefficient pressure-Poisson solver is needed.
\citet{rudman1998} established the importance of using compatible mass and momentum fluxes, and subsequent formulations have developed this consistency for staggered-grid VOF methods~\citep{vaudor2017,arrufat2021}.
Inconsistent transport can introduce non-physical velocities near the interface.
Such errors are particularly relevant to particle tracking because the interpolated fluid velocity directly enters the particle force.

Modelling particle motion thus requires a description of the forces acting on individual particles and can involve relatively complex interactions~\citep{yao2022,yao2021,vowinckel2021}. 
Lagrangian point-particle methods offer a computationally economical description of dilute suspensions by modelling the hydrodynamic forces rather than resolving the flow around every particle.
Their applicability depends on the particle size and density, as well as on the time and length scales of the carrier flow.
The small-sphere equation of~\citet{maxey1983} includes contributions from fluid acceleration, added mass and viscous history in addition to drag and buoyancy.
For heavy particles, the large particle-to-fluid density ratio can justify neglecting fluid-acceleration and added-mass contributions relative to drag~\citep{maxey1983}.
For light particles in an accelerating liquid, the omitted force contributions require closer consideration.
At the free surface, a further distinction arises between tracking a point across a phase boundary and describing the changing immersion of a finite floating object~\citep{calvert2021}.
These regimes place different demands on the particle model even when they share the same underlying fluid solver.

In this paper, we present a computational framework: Discrete Operator-based Parallel Algorithm for Modelling Incompressible Navier--Stokes Equations (\textit{Dopamine})\footnote{The source code is available at \url{https://github.com/AkshayPatil1994/dopamine-fdm}.}, which combines the above-discussed geometric VOF transport, consistent mass--momentum advection and a variable-density pressure projection with one-way-coupled Lagrangian particle tracking using low-dissipation numerical methods and state-of-the-art sub-grid-scale turbulence models within the large-eddy simulation approach.
The particle formulation uses drag and buoyancy, with options for confinement to the liquid or free-surface crossing followed by ballistic motion in the gas.
We compare the proposed numerical framework against analytical, experimental, and numerical benchmarks to demonstrate its efficacy in reproducing the essential physics of coastal ocean waves and particle transport.
In the following sections, we present the governing equations and numerical methods, and discuss the numerical and experimental comparisons.

\section{Problem formulation}
\label{sec:problem-formulation}
\subsection{Two-fluid governing equations}
Dopamine is built on the single-phase solver originally developed by~\citet{lozanoduran2019} and further extended by~\citet{patilfringer2022,patilgarciasanchez2025} and solves the incompressible Navier--Stokes equations for immiscible liquid and gas phases on a staggered Cartesian grid, with an optional Boussinesq thermal-buoyancy term.
The coordinate system $x_i$ is aligned such that $x_1$, $x_2$, and $x_3$ are streamwise, vertical and spanwise, with gravity acting along $-x_2$; all vectors and tensors are written in index notation with respect to this basis, and the vertical elevation is $x_2$.
The conservative one-fluid equations, including the continuum surface-tension force~\citep{brackbill1992}, are given by
\begin{align}
  \partial_t\rho+\partial_j(\rho u_j)&=0,\\
  \partial_t(\rho u_i)+\partial_j(\rho u_i u_j)&=-\partial_i p+\rho g_i
  +\partial_j\left[(\mu+\mu_t)(\partial_j u_i+\partial_i u_j)\right]
  +\sigma\kappa\partial_i C+f_{B,i},
  \label{eq:momentum}\\
  \partial_i u_i&=0,\\
  \partial_t C+\partial_j(Cu_j)&=0.
  \label{eq:c}
\end{align}
Repeated indices imply summation, and $\delta_{ij}$ is the Kronecker delta.
Here, $C$ is the liquid fraction, $u_i$ is the velocity, $p$ is the pressure, and $g_i=-g\delta_{i2}$ is the gravitational acceleration.
Surface tension and curvature are denoted by $\sigma$ and $\kappa$.
Density and dynamic viscosity are
\begin{equation}
  \rho=\rho_g+(\rho_l-\rho_g)C,\qquad
  \mu=\mu_g+(\mu_l-\mu_g)C,
\end{equation}
where $l$ and $g$ denote liquid and gas, and the SGS dynamic viscosity is $\mu_t=\rho\nu_t$.

When temperature stratification is enabled, the Boussinesq force per unit volume is given by
\begin{equation}
  f_{B,i}=-C_B\rho_l\beta_T(T-T_{\mathrm{ref}})g_i.
  \label{eq:boussinesq-force}
\end{equation}
Here $T$ is temperature, $T_{\mathrm{ref}}$ is its reference value, and $\beta_T$ is the thermal expansion coefficient.
Since gravity acts along $-x_2$, a positive temperature anomaly produces an upward force.
The factor $C_B$ weights this force by the liquid fraction; at a vertical-velocity face it is the arithmetic mean of the two adjacent cell fractions.
The force vanishes in the pure gas phase.
Temperature-dependent density perturbations are included in this acceleration/force term, while phase densities are used in mass transport, inertia, and the pressure operator.
The Boussinesq approximation assumes $|\beta_T(T-T_{\mathrm{ref}})|\ll1$ within the liquid; the liquid--gas density contrast is retained independently by the two-fluid formulation.
Without temperature stratification, $f_{B,i}=0$.

Temperature satisfies the transport equation given by
\begin{equation}
  \partial_t T+u_j\partial_j T
  =\partial_j\left[(\alpha(C)+\alpha_t)\partial_j T\right],
  \qquad
  \alpha(C)=\frac{C\nu_l+(1-C)\nu_g}{Pr},
  \label{eq:temperature}
\end{equation}

\noindent here, $Pr$ is the molecular Prandtl number, and $\alpha_t$ is the SGS thermal diffusivity.
The turbulent sub-grid-scales (SGS) are modelled by closure models that provide the SGS viscosity $\mu_t$ and diffusivity $\alpha_t$.
Two established sub-grid-scale models are included: the Vreman model~\citep{vreman2004} and the anisotropic minimum-dissipation (AMD) model~\citep{rozema2015}.
For Vreman LES, $\alpha_t=\nu_t/Pr_t$, where $Pr_t$ is the turbulent Prandtl number and $\nu_t$ is the SGS kinematic viscosity.
The AMD option instead evaluates its scalar diffusivity from the resolved velocity and temperature gradients~\citep{abkar2016}, with the viscosity-based expression used in its immersed-boundary fallback band.
The implemented temperature field extends through both the fluid phases, although its buoyancy acts only in the liquid phase.
It is advanced once per fluid time step with the final velocity using a third-order strong-stability-preserving Runge--Kutta (SSP-RK3) scheme~\citep{shu1988,gottlieb1998}; molecular diffusivities are harmonically averaged at cell faces.
The scalar advection includes a discrete divergence correction to preserve uniform temperature when the projected velocity has a non-zero divergence residual.
In the two-fluid update, momentum and particles use the temperature available before this scalar step, so the thermal coupling is lagged; the SSP-RK3 scalar substep alone does not establish third-order accuracy of the coupled thermal flow.

\subsection{Interface and momentum transport}
Pressure and liquid fraction, being scalar quantities, are stored at cell centres while each velocity component lies at the corresponding face, as in the marker-and-cell arrangement of~\citet{harlow1965}.
Mixed cells contain a piecewise-linear interface (PLIC), while geometric volume tracking and construction are obtained using the algorithm described by~\citet{rider1998}.
Finally, the interface normals are obtained from a Youngs-type reconstruction or the centred-column height-function option.

Directionally split geometric fluxes advance the scalar field $C$ based on the methodology proposed by~\citet{weymouth2010}.
Each sweep includes the compression correction for directional divergence, and the sweep order alternates each time step.
The transporting velocity is held fixed during an advection half-step.
Sub-cycling targets a sub-step Courant number no larger than 0.12 by default.
At liquid-to-gas density ratios of 2000 or greater, the default pseudo-time momentum integrator switches to forward Euler and the sub-step Courant target is reduced to at most 0.03.

The momentum update uses the mass fluxes that advance density, following~\citet{rudman1998},~\citet{vaudor2017} and~\citet{arrufat2021}.
On a staggered momentum control volume, the face mass flux is constructed from the neighbouring cell fluxes and the split-scheme correction.
WENO5-Z reconstructs the transported velocity~\citep{borges2008}, blending towards first-order upwinding where local mass turnover requires it near the interface.
Within each directional sweep, the default momentum update uses pseudo-time SSP-RK3 with fixed geometric mass fluxes and a linearly varying control-volume density.
The complete two-fluid step comprises an advection half-step, a projected three-stage force update at fixed density, and a second advection half-step followed by projection.
The order of these split operations, the transporting-velocity approximation and the pressure tolerance must therefore be considered when assessing temporal convergence; the order of an individual Runge--Kutta substep does not establish that of the complete solver.

\subsection{Pressure, viscous stresses and surface tension}
Incompressibility is enforced by projecting the provisional velocity onto a divergence-free field. The pressure correction satisfies the variable-coefficient Poisson equation
\begin{equation}
  \partial_i\left(\frac{1}{\rho_{\mathrm I}}\partial_i p\right)
  =\frac{1}{\Delta t}\partial_i u_i^*,
\end{equation}
where $u_i^*$ denotes the provisional velocity and $\rho_{\mathrm I}$ is the density entering the pressure operator.
By default, $\rho_{\mathrm I}$ is the geometric face density obtained from the PLIC liquid volumes in the two adjoining half-cells.
An alternative inertia treatment uses the arithmetic control-volume mass transported by the momentum scheme in the pressure operator and force accelerations, while retaining geometric densities in the force numerators.
This alternative is intended to suppress excessive gas-side acceleration during wave breaking and complex interface evolution.
The implementation subtracts a still-water hydrostatic reference, so the stored pressure is a reduced pressure rather than the total pressure appearing in Eq.~\eqref{eq:momentum}.

The pressure increment in the force update is solved by preconditioned conjugate gradients (PCG), with a default tolerance of 0.2 on the residual norm relative to its initial value and a maximum of 30 iterations.
The norm is weighted by active-cell volume.
Separate projections after the advection half-steps use a relative tolerance of $10^{-8}$ with a default cap of three iterations, while the transporting-velocity projection has its own absolute residual floor.
For wall-bounded vertical configurations, a mask-aware multigrid method supplies the default preconditioner; a constant-coefficient fast Poisson solver is also available on CPUs only.

The total molecular-plus-SGS viscosity required for the shear-stress discretisation is interpolated from the four neighbouring cell centres to each shear edge using a weighted harmonic mean.
The surface-tension force is discretised using the balanced-force formulation with height-function curvature~\citep{popinet2009}.
Where the interface geometry does not admit a valid height-function column, curvature is estimated from neighbouring valid values and set to zero only when no such values are available; the number of these fallback evaluations is recorded.
The time step is constrained by both capillary and gravity-wave restrictions to ensure numerical fidelity is preserved when integrating the governing equations in time.

At immersed bodies, velocity boundary conditions are imposed through the ghost-cell construction of~\citet{tseng2003}, whereas the pressure operator and VOF transport represent impermeable boundaries by closed staircase faces.
The geometric accuracy of the pressure and volume-fraction boundary treatment is therefore lower than that of the velocity construction.
Optional equilibrium wall models replace the viscous shear flux by a smooth- or rough-wall log-law stress at flat walls and immersed staircase faces.
This replacement is restricted to locally single-phase stencils; near the liquid--gas interface the underlying viscous treatment is retained.
The interface may intersect an immersed body, but the formulation has no contact-angle condition or moving-contact-line model and therefore does not resolve wetting dynamics or capillary attachment to the body.

\subsection{Wave generation and absorption}
Wave generation is achieved through prescribed inlet elevation and velocity, together with relaxation zones for generation and absorption~\citep{jacobsen2012}.
The prescribed fields may be obtained from linear wave theory, a Rienecker--Fenton stream-function solution~\citep{rienecker1981}, a JONSWAP component sum~\citep{hasselmann1973}, or a Fourier decomposition of a measured elevation record.
For irregular waves, the velocity field uses finite-depth dispersion and Wheeler stretching~\citep{wheeler1970}; the record-driven formulation retains a linear superposition and therefore does not reconstruct nonlinear bound harmonics.
Within the relaxation zones, velocity and liquid fraction are blended towards the prescribed wave field or still water using the relaxation formulation of~\citet{jacobsen2012}.

Because this blending can add or remove liquid, volume conservation is evaluated after accounting for both boundary fluxes and relaxation-zone contributions:
\begin{equation}
  R_V(t)=V_l(t)-V_l(0)-\Delta V_{\mathrm{boundary}}(t)
  -\Delta V_{\mathrm{relaxation}}(t),
  \label{eq:ledger}
\end{equation}
where $V_l=\int_{\Omega_f}C\,\dd V$ is the liquid volume in the fluid region, and the boundary and relaxation terms are accumulated separately.
The inlet balances the instantaneous water-column flux through a gas return flow.
For stream-function waves, the prescribed mean flow can enforce either zero mean volume flux or zero mean Eulerian velocity; these choices do not remove the need to measure the simulated water-column mean current when interpreting particle drift.

\subsection{Lagrangian particle model}
The motion of a particle with position $x_{p,i}$ and velocity $v_{p,i}$ is driven by the local carrier velocity, obtained by trilinear interpolation of each staggered velocity component to give $u_{f,i}=u_i(x_{p,j},t)$.
The particle dynamics are described by a drag--buoyancy reduction of the small-sphere equation~\citep{maxey1983},
\begin{align}
  \frac{\dd x_{p,i}}{\dd t}&=v_{p,i},\\
  \frac{\dd v_{p,i}}{\dd t}&=
  \frac{u_{f,i}-v_{p,i}}{\tau_{\mathrm{eff}}}
  -g\left(1-\frac{\rho_{f,B}}{\rho_p}\right)\delta_{i2}
  +a_{\mathrm{opt},i},
  \label{eq:particle}
\end{align}
with response times and particle Reynolds number defined by
\begin{gather}
  \tau_p=\frac{\rho_p d_p^2}{18\rho_f\nu_l},\qquad
  \tau_{\mathrm{eff}}=\frac{\tau_p}{1+0.15\Rey_p^{0.687}},\\
  \Rey_p=\frac{\sqrt{(u_{f,j}-v_{p,j})(u_{f,j}-v_{p,j})}\,d_p}{\nu_l},
\end{gather}
where $d_p$ and $\rho_p$ are particle diameter and density, $\nu_l$ is liquid kinematic viscosity, and $\rho_f=\rho_l$ in the liquid.
The density entering particle buoyancy is $\rho_{f,B}=\rho_f$ unless temperature stratification is enabled together with particle thermal coupling.
With this coupling the particle buoyancy is evaluated using
\begin{equation}
  \rho_{f,B}=\rho_f\left[1-\beta_T\bigl(T(x_{p,j})-T_{\mathrm{ref}}\bigr)\right].
  \label{eq:particle-thermal-buoyancy}
\end{equation}
The temperature in this expression is interpolated to the particle position, while the reference fluid density is retained in the drag relaxation time and optional added-mass coefficient.
The coupling is one-way, with no particle feedback on carrier momentum or temperature.
The default drag correction uses the Schiller--Naumann form~\citep{schiller1935,clift1978,marchioli2008}, with nominal applicability $\Rey_p\lesssim1000$; a linear Stokes-drag option sets the correction factor to unity.
The point-particle approximation additionally requires $d_p\ll\Delta$, where $\Delta$ is the local grid scale, and a diameter small relative to the sampled flow scales.

Fluid-acceleration effects can be included through two alternative formulations.
The first retains the drag--buoyancy equation and sets $a_{\mathrm{opt},i}=a_{\mathrm{AM},i}$, where
\begin{equation}
  a_{\mathrm{AM},i}=\frac{1}{2}\frac{\rho_f}{\rho_p}
  \frac{u_{f,i}^{\mathrm{new}}(x_{p,j})-u_{f,i}^{\mathrm{ref}}(x_{p,j})}{\Delta t},
\end{equation}
both velocity fields are sampled at the same particle position.
This local time-difference correction leaves particle inertia unchanged and omits convective fluid acceleration.
The second formulation replaces the acceleration equation by the reduced Maxey--Riley model~\citep{maxey1983},
\begin{align}
  \frac{\dd v_{p,i}}{\dd t}
  &=\frac{u_{f,i}-v_{p,i}}{\tau_{\mathrm{MR}}}
    +\beta_{\mathrm{MR}}\frac{D u_{f,i}}{D t}
    +(1-\beta_{\mathrm{MR}})g_i,\label{eq:particle-mr}\\
  \beta_{\mathrm{MR}}&=\frac{3\rho_f}{\rho_f+2\rho_p},\qquad
  \tau_{\mathrm{MR}}=\left(1+\frac{\rho_f}{2\rho_p}\right)\tau_{\mathrm{eff}}.
\end{align}
Here $D/Dt=\partial_t+u_j\partial_j$ denotes the carrier-fluid material derivative, evaluated at the particle position.
Its numerical approximation combines the final-minus-reference velocity difference divided by $\Delta t$ with centred differences of the spatially interpolated velocity for the convective term.
The model accounts for added inertia and fluid acceleration but neglects history, lift, rotation and finite-size corrections.

Particle advancement follows the final fluid-velocity projection and spans one fluid time step.
For inertial particles, the final carrier field is sampled at $x^n_{p,i}$, and the relaxation time is computed from the slip relative to $v^n_{p,i}$.
These quantities and the non-drag acceleration $a_i$, comprising buoyancy-reduced gravity and any optional acceleration, are held fixed during integration.
For the drag--buoyancy formulation, direct integration of Eq.~\eqref{eq:particle} then gives
\begin{align}
  v_{\infty,i}&=u_{f,i}+\tau_{\mathrm{eff}}a_i,\\
  v^{n+1}_{p,i}&=v_{\infty,i}
  +(v^n_{p,i}-v_{\infty,i})
  \exp(-\Delta t/\tau_{\mathrm{eff}}),\\
  x^{n+1}_{p,i}&=x^n_{p,i}+v_{\infty,i}\Delta t
  +\tau_{\mathrm{eff}}(v^n_{p,i}-v_{\infty,i})
  [1-\exp(-\Delta t/\tau_{\mathrm{eff}})].
\end{align}
For Eq.~\eqref{eq:particle-mr}, the same expressions apply with $\tau_{\mathrm{MR}}$ replacing $\tau_{\mathrm{eff}}$ and $a_i=\beta_{\mathrm{MR}}D u_{f,i}/Dt+(1-\beta_{\mathrm{MR}})g_i$.
The exponential update is exact for the frozen-coefficient problem and removes the explicit stability restriction associated with the relaxation time.
However, temporal error may arise from neglecting changes in the sampled velocity, drag correction and non-drag acceleration during the step.

Tracer positions are advanced using the classical three-stage RK3 method, with velocity evaluated at the stage positions and times $t_n$, $t_n+\Delta t/2$ and $t_{n+1}$.
The endpoint stages use the corresponding stored velocity fields, while the midpoint field is reconstructed by quadratic Lagrange interpolation through $t_{n-1}$, $t_n$ and $t_{n+1}$.
For equal time steps,
\begin{equation}
  \widetilde u_i(\boldsymbol x,t_n+\Delta t/2)
  =-\tfrac18u_i^{n-1}(\boldsymbol x)
   +\tfrac34u_i^n(\boldsymbol x)
   +\tfrac38u_i^{n+1}(\boldsymbol x).
\end{equation}
The interpolation weights account for unequal consecutive time steps.
With smooth velocity fields, sufficiently accurate stored solutions, consistent initialisation and bounded ratios of consecutive step lengths, the tracer position update is formally third-order accurate in time.
The first step uses linear interpolation between the available fields until a third time level has been stored.

Two implementation details qualify the third-order temporal accuracy for wave--tracer calculations.
First, the velocity history is omitted from particle checkpoints and reconstructed after restart, so restarted trajectories need not coincide exactly with an uninterrupted calculation.
Furthermore, the two-fluid solver initialises this history from the velocity after the first advection half-step rather than from the true start-of-step field.

Optional SGS dispersion is represented by a simplified Langevin velocity fluctuation, following the general stochastic modelling approach discussed by~\citet{shotorban2006}.
The fluctuation is advanced using Euler--Maruyama integration~\citep{higham2001} and held fixed during the particle update.
For inertial particles, Brownian motion can additionally be represented by a random displacement with Stokes--Einstein diffusivity~\citep{einstein1905}.
The deterministic accuracy statements above do not apply to these stochastic updates or to discrete collision and surface-crossing events.
Particles may be initialised at rest or at the local fluid velocity; immersed-body collisions support reflection and absorption, with optional speed-dependent deposition.
Particle IDs are retained during MPI migration, and particle state can be restored from checkpoints.
The checkpoints do not preserve the random-number state or the SGS velocity fluctuation, so stochastic restarts are not exact continuations.

The default boundary treatment confines particles to the liquid by rejecting trial steps that end at interpolated $C<0.5$ and reflecting the outward normal velocity.
When free-surface crossing is enabled, an inertial particle instead undergoes a split update at the interface.
The crossing point, defined by $C=0.5$, is located by 20 bisections of the straight trial segment; the liquid update is then repeated over the corresponding fraction of the time step, and the remainder is integrated ballistically.
Both phase classification and the segment search use the final liquid-fraction field, held fixed during the particle step.
Identifying the segment fraction with the elapsed-time fraction therefore approximates both particle acceleration and interface motion; the 20 bisections refine the segment intersection at negligible computational cost.

In the gas phase, the update follows an exact ballistic solution that does not interact with the gas-phase velocity field,
\begin{equation}
  x_{p,i}(t+s)=x_{p,i}(t)+s\,v_{p,i}(t)
  -\tfrac12gs^2\delta_{i2},\qquad
  v_{p,i}(t+s)=v_{p,i}(t)-gs\,\delta_{i2}.
\end{equation}
This gravity-only description neglects gas drag and carrier-gas velocity, while the crossing rule excludes capillary retention and partial submergence.
Liquid forces are restored when a subsequent step begins at $C\ge0.5$, giving a coarser temporal resolution of re-entry than of exit.
The model therefore describes the crossing of a point particle through the interface, with no representation of the immersion-dependent buoyancy of finite floating objects such as those studied by~\citet{calvert2021}.

\section{Results and discussion}
\label{sec:results-and-discussion}

The implementation is first verified using analytical cases and thereafter validated against established numerical and experimental data.
For the quasi-two-dimensional simulations, the quasi-two-dimensional simulations use periodic boundary conditions along the spanwise direction ($x_3$), with a small number of cells specified for each case.
Unless dimensional units are stated, each case retains its own nondimensional convention.

\begin{table}[htbp]
  \centering
  \caption{Analytical verification cases for the equilibrium, wave, density-driven and particle components.
  Here, $N$ denotes the number of grid points, $a$ denotes the wave amplitude, and the error metrics for each case use different reference values.}
  \label{tab:components}
  \small
  \begin{tabular}{@{}>{\raggedright\arraybackslash}p{0.34\linewidth}>{\raggedright\arraybackslash}p{0.28\linewidth}>{\raggedright\arraybackslash}p{0.28\linewidth}@{}}
    \toprule
    Case & Coarse ($N = 32$) & Fine ($N = 64$)\\
    \midrule
    Hydrostatic rest & Zero to round-off & Zero to round-off\\
    Finite-depth waves & Frequency error $\le2.06\%$ & Frequency error $\le1.00\%$\\
    Rayleigh--Taylor & Growth error $-9.1$ to $-7.9\%$ & Growth error $-8.1$ to $-7.6\%$\\
    Periodic lock exchange & $\mathrm{Fr}=0.490$ & $\mathrm{Fr}=0.495$\\
    Standing-wave tracers & RMS error$/a=0.167$ & RMS error$/a=0.104$\\
    Standing-wave inertial particles &RMS error$/a=0.088$ & RMS error$/a=0.054$\\
    Progressive-wave tracers & Mean drift/reference $=0.724$ & Mean drift/reference $=0.851$\\
    \bottomrule
  \end{tabular}
\end{table}

\subsection{Hydrostatic and capillary equilibrium verification}
Hydrostatic equilibrium tests the discrete balance between pressure and gravity across the phase-density contrast.
For density ratios 1.2, 10 and 1000, the column remains at rest to round-off as detailed in Table~\ref{tab:components}.
The reported pressure excludes the hydrostatic contribution, so the nearly vanishing difference between phases refers to this reduced variable rather than the total pressure head.

\begin{figure}[htbp]
  \centering
  \includegraphics[width=\linewidth]{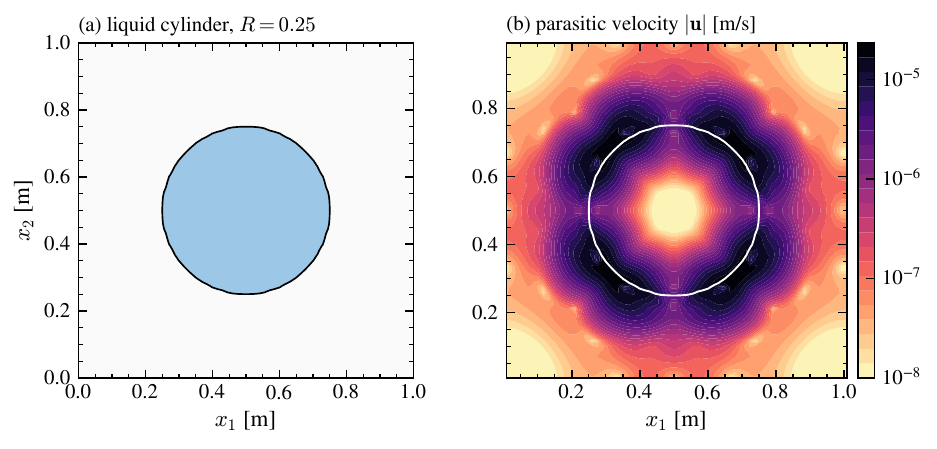}
  \caption{Static liquid cylinder at density ratio 1000 and $R=0.25$ m: (a) liquid fraction and $C=0.5$ contour; (b) parasitic velocity magnitude on a logarithmic colour scale.
  The panels show the final mid-plane slice at $N=64$.}
  \label{fig:static-drop}
\end{figure}

The static-cylinder cases span radii $R=0.2$, 0.25 and 0.3 m at surface tension $\sigma=1\,\mathrm{N\,m^{-1}}$, with a further case at $R=0.25$ m and $\sigma=4\,\mathrm{N\,m^{-1}}$.
With gravity disabled, density ratio 1000 and equal phase kinematic viscosities of $10^{-3}\,\mathrm{m^2\,s^{-1}}$, the equilibrium pressure jump is $\sigma/R$.
Relative to this analytical value, the pressure errors decrease from 0.53--1.27\% at $N=32$ to 0.15--0.29\% at $N=64$.
For $R=0.25$ m and $\sigma=1\,\mathrm{N\,m^{-1}}$, the pressure jump decreases from 4.0336 to 4.0078 Pa, compared with the exact 4 Pa, while the maximum speed over the final 20 diagnostic samples decreases from $1.1\times10^{-4}$ to $1.8\times10^{-5}\,\mathrm{m\,s^{-1}}$.
The reduction in parasitic motion accompanies the improvement in pressure balance. Fig.~\ref{fig:static-drop} shows the interface and instantaneous velocity magnitude in the $N=64$ mid-plane slice, while the reported maximum is evaluated over the final sampling window.
For the three-dimensional sphere, the pressure jumps are 8.083 and 8.030 at $N=24$ and 32, compared with $2\sigma/R=8$ in the same units.
Agreement of the capillary-wave frequency with linear dispersion to within 1\% further assesses the dynamic response of the interface-force discretisation~\citep{prosperetti1981}.

\subsection{Particle-response verification}
\label{sec:particle-response-verification}
\begin{figure}[htbp]
  \centering
  \includegraphics[width=\linewidth]{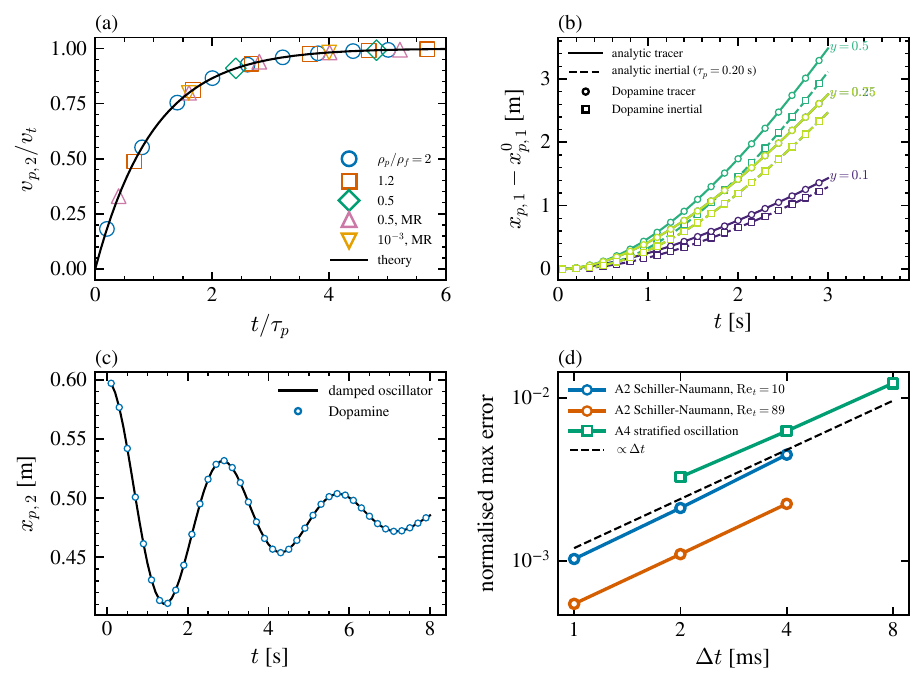}
  \caption{Particle-response verification: (a) normalised Stokes settling and rise, with MR denoting the quiescent limit of the reduced Maxey--Riley formulation; (b) tracer and inertial-particle displacement in start-up Poiseuille flow; (c) stratified-particle displacement and the damped-oscillator reference with fitted equilibrium height; (d) normalised maximum errors under time-step refinement for nonlinear drag and stratified buoyancy.
  Symbols show Dopamine results and lines show analytical or independently integrated references.
  In panel (a), open circles, squares, diamonds, upward triangles and downward triangles distinguish the five cases listed in the legend; the relaxation time includes added inertia where enabled.}
  \label{fig:particle-dynamics}
\end{figure}

Particle integration is first assessed in quiescent liquid, isolating the drag and buoyancy response from free-surface motion.
The settling and rising trajectories in Fig.~\ref{fig:particle-dynamics}(a) are compared with the analytical exponential response under linear Stokes drag.
The case uses liquid density $1000\,\mathrm{kg\,m^{-3}}$, kinematic viscosity $10^{-4}\,\mathrm{m^2\,s^{-1}}$ and particle diameter 6.7 mm.
The particle-to-fluid density ratios are 2, 1.2 and 0.5 without added mass, and with ratios of 0.5 and $10^{-3}$ using the reduced Maxey--Riley formulation (added mass included).
For the latter cases, the reference relaxation time is $(\rho_p+\rho_f/2)d_p^2/(18\mu)$, while the terminal velocity retains its Stokes value.
The normalised responses follow $1-\exp(-t/\tau)$, as expected from the exact integration of the constant-coefficient relaxation equation.
In this quiescent limit, the material-acceleration term in Eq.~\eqref{eq:particle-mr} vanishes, so the comparison verifies added inertia and buoyancy without testing the discretisation of fluid acceleration in an evolving carrier.

The start-up Poiseuille case tests interpolation and particle response in an evolving carrier flow.
The channel height is 1 m, kinematic viscosity is $0.05\,\mathrm{m^2\,s^{-1}}$, and the imposed streamwise acceleration is $1\,\mathrm{m\,s^{-2}}$.
Tracer and inertial particles begin at four heights, with an inertial relaxation time of approximately 0.20 s.
The tracer reference integrates the analytical velocity series; the inertial reference independently integrates the drag equation driven by that series.
At $N=64$, the maximum displacement errors, normalised by the maximum reference displacement over 3 s, are at most 0.08\% for tracers and 0.07\% for inertial particles.
The trajectories in Fig.~\ref{fig:particle-dynamics}(b) accordingly show close agreement with the reference, although non-monotonic error changes between the two meshes preclude assigning a spatial convergence order.

A stably stratified, initially quiescent carrier supplies a further test of position-dependent buoyancy.
With Stokes drag, the reference displacement satisfies a damped-oscillator equation with the relaxation time and restoring coefficient specified by the case parameters.
The comparison in Fig.~\ref{fig:particle-dynamics}(c) uses a fitted equilibrium height to account for the temperature-profile origin, while retaining the prescribed frequency and relaxation time.
The time-step refinement in panel (d) includes this case and finite-Reynolds-number Schiller--Naumann settling, for which reference trajectories are obtained by independent ODE integration.
The plotted errors decrease approximately linearly with the time step, consistent with freezing the changing buoyancy or drag coefficient over a particle step.

\subsection{Particle-interface interaction verification}
\begin{figure}[htbp]
  \centering
  \includegraphics[width=\linewidth]{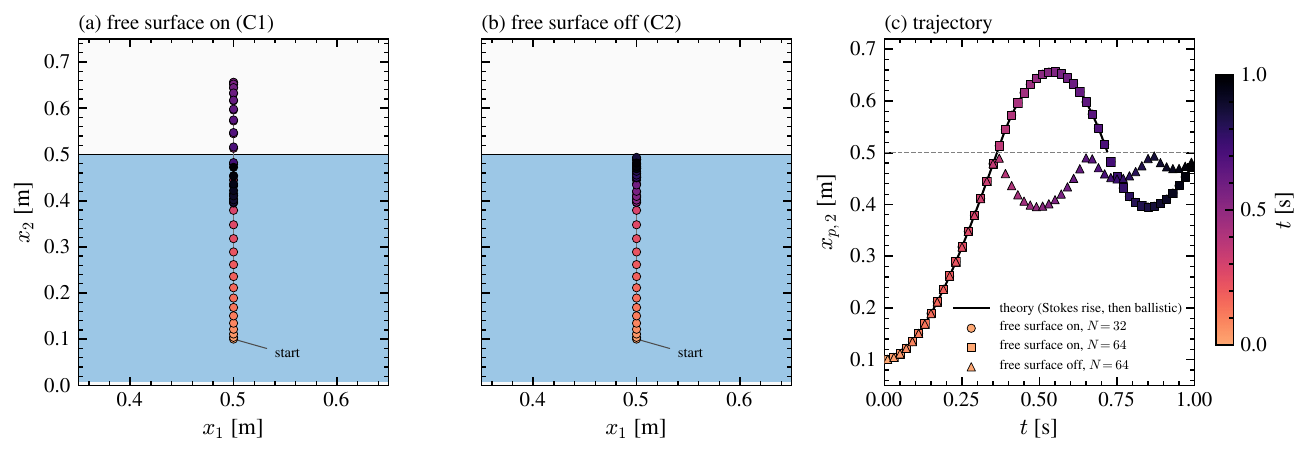}
  \caption{Buoyant-particle rise in static tank with a horizontal interface: (a) trajectory with surface crossing enabled; (b) trajectory with surface crossing disabled such that the particles are confined to the liquid phase; (c) vertical position compared with Stokes rise followed by ballistic flight.
  Symbol colour denotes time.
  The analytical flight curve ends at its first return to the interface; subsequent liquid re-entry is not covered by that reference.}
  \label{fig:surface-exit}
\end{figure}

The transition from liquid-phase rise to airborne motion is verified against a piecewise analytical solution for the first rise-and-flight event beneath a flat interface.
A particle of diameter 0.02 m and density $500\,\mathrm{kg\,m^{-3}}$ starts from rest at $x_2=0.1$~m below the interface, which is located at $x_2=0.5$~m, as sketched in Fig.~\ref{fig:surface-exit}(a).
The liquid has a density of $1000\,\mathrm{kg\,m^{-3}}$, kinematic viscosity of $5\times10^{-5}\,\mathrm{m^2\,s^{-1}}$, and linear Stokes drag is imposed on the particles.
With this setup, the particle exits the interface with an exit velocity of $1.7527\,\mathrm{m\,s^{-1}}$ at $0.362$~s, and a ballistic apex of $x_2=0.656$~m is reached.
Both $N=32$ and 64 reproduce the analytical apex, whereas disabling surface crossing confines the maximum height to 0.5 m (Fig.~\ref{fig:surface-exit}(b)).
The numerical and analytical trajectories agree for both the confined particle and the particle that exits the liquid, follows a ballistic trajectory and re-enters the liquid.

In a second flat-tank check, 5 mm particles of density $500\,\mathrm{kg\,m^{-3}}$ rise through water and leave the surface at about $0.29\,\mathrm{m\,s^{-1}}$.
The fitted airborne deceleration of $9.81\,\mathrm{m\,s^{-2}}$ and apex approximately 4.3 mm above the surface agree with the ballistic prediction $v_e^2/(2g)$ for exit speed $v_e$.
For water with $\nu_l=10^{-6}\,\mathrm{m^2\,s^{-1}}$, however, the exit speed corresponds to $\Rey_p\approx1450$, exceeding the nominal range of the drag law.
This comparison therefore verifies airborne integration conditional on the supplied exit velocity, without establishing the accuracy of the preceding rise or physical escape through the interface.

\subsection{Interface advection validation}
Interface-transport accuracy is assessed by prescribed-velocity return tests, in which the final volume-fraction field is compared with its initial distribution (Table~\ref{tab:transport}).
The two-dimensional vortex and three-dimensional deformation cases follow~\citet{leveque1996} and~\citet{enright2002}, respectively.
The vortex starts with a disk of radius 0.15 in the unit box and has cycle duration $T=8$; the three-dimensional case starts with a sphere of radius 0.15 and has $T=3$.
For the vortex, successive grid refinement reduces the return error to $6.7\times10^{-5}$ at $N=128$.
For the three-dimensional case, the error decreases from $7.9\times10^{-3}$ to $5.4\times10^{-3}$ between $N=32$ and $N=48$.

\begin{figure}[htbp]
  \centering
  \includegraphics[width=\linewidth]{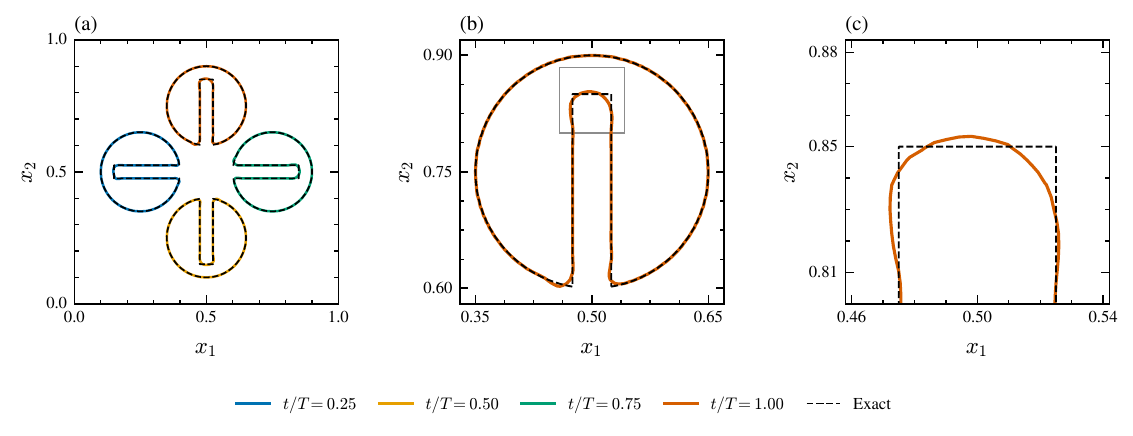}
  \caption{Zalesak slotted disk comparison using a $256\times256$ grid over the domain: (a) numerical $C=0.5$ contours during one revolution at four distinct times; (b) numerical solution (solid line) after one full rotation against the exact outline (dashed line); (c) zoomed view of panel (b) near the square notch.
  Solid colours denote the rotation times; black dashes denote the exact interface.}
  \label{fig:zalesak}
\end{figure}

\begin{table}[htbp]
  \centering
  \caption{Interface advection error for the two- and three-dimensional return tests as a function of grid resolution. Here, the error is defined as $E_C=\int|C(T)-C(0)|\,\dd V$.}
  \label{tab:transport}
  \resizebox{\linewidth}{!}{
    \begin{tabular}{llll}
      \toprule
      Case & Grid $N$ & $E_C$ & Comment\\
      \midrule
      Reversed disk vortex & 32 / 64 / 128 & $8.5\times10^{-3}$ / $9.8\times10^{-4}$ / $6.7\times10^{-5}$ & Unit box\\
      3-D sphere deformation & 32 / 48 & $7.9\times10^{-3}$ / $5.4\times10^{-3}$ & Thin sheet\\
      Slotted disk rotation & 100 / 256 & $2.0\times10^{-3}$ / $6.5\times10^{-4}$ & One revolution\\
      \bottomrule
    \end{tabular}
  }
\end{table}
The Zalesak disk~\citep{zalesak1979} has radius 0.15, centre $(x_1,x_2)=(0.5,0.75)$ and slot width 0.05 in the unit box and is rotated rigidly by a prescribed velocity field.
The comparison of the numerical solution with the exact pure-advection solution is shown in Fig.~\ref{fig:zalesak}, which shows progressive rounding of the slot corners, as also discussed by~\citet{pan2024}.
Although liquid fractions remain bounded without numerical clipping and volume drift is accurate to round off ($7\times10^{-15}$), the enlarged slot contour retains a visible shape error as expected.

\subsection{Lock-exchange validation}
\label{sec:lock-exchange}
\begin{figure}[htbp]
  \centering
  \includegraphics[width=\linewidth]{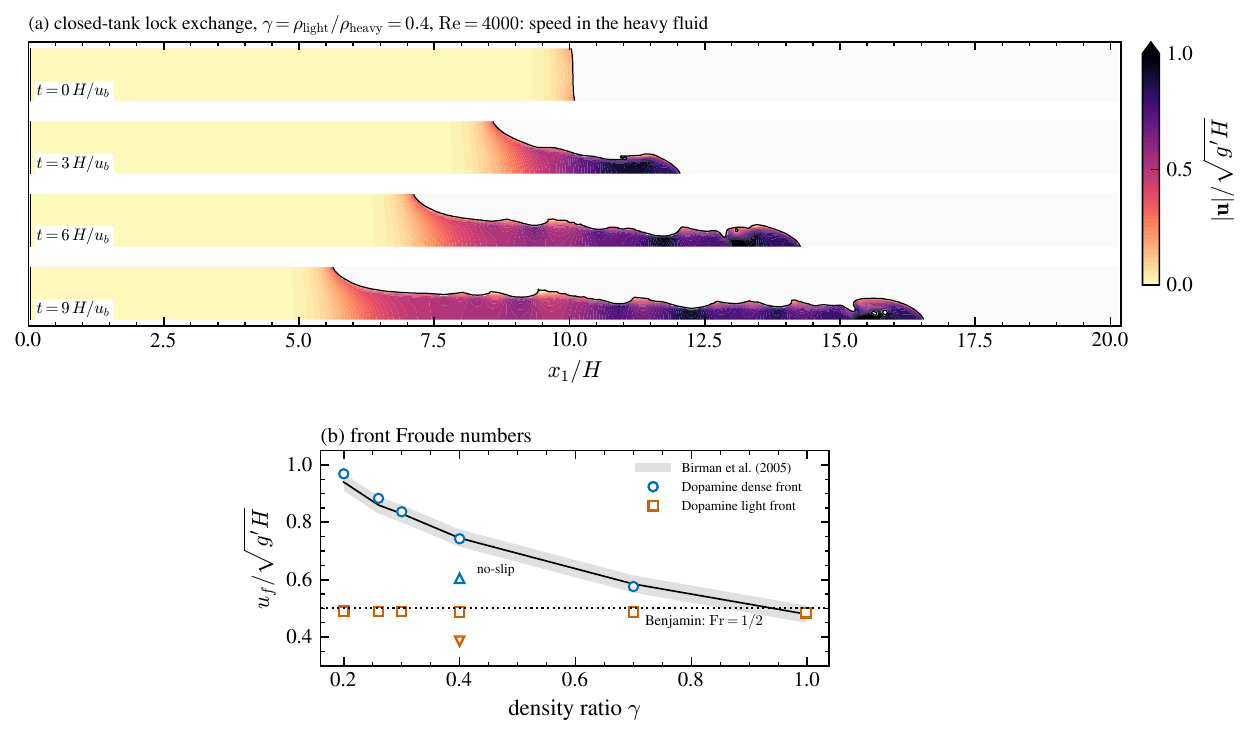}
  \caption{Closed-tank lock exchange: (a) heavy-fluid speed and $C=0.5$ contours at four times for $\gamma=0.4$; (b) dense- and light-front Froude numbers across the density-ratio sweep. The dotted line denotes $\mathrm{Fr}=1/2$; open triangles denote the additional no-slip case.}
  \label{fig:lock-exchange}
\end{figure}

Density-driven interface motion is assessed using the classic lock-exchange problem, which tests the two-fluid momentum balance in the absence of imposed wave forcing.
The initial condition consists of a full-depth density interface dividing a tank of length $20H$ and height $H$ as shown in the top panel in Fig.~\ref{fig:lock-exchange}.
Immersed solid slabs close the ends inside the periodic computational domain with free-slip boundaries along the horizontal directions.
Using the density ratio defined as $\gamma=\rho_{\mathrm{light}}/\rho_{\mathrm{heavy}}$, the reduced gravity then becomes $g'=g(1-\gamma)$ and the front velocity is thus obtained as $u_b=\sqrt{g'H}$.
The front velocity and domain height-based Reynolds number are $\Rey_{b} \equiv u_bH/\nu=4000$, with equal kinematic viscosities and 32 cells per tank height, for a $\gamma$ ranging from $0.2$ to $0.998$.
Front speeds are fitted to the near-wall $C=0.5$ crossings over $3\le tu_b/H\le9$.
Fig.~\ref{fig:lock-exchange}(a) shows the two counter-propagating currents and the speed within the heavy fluid.

The light-front Froude numbers $U_{\mathrm{fr}}/u_b$ range from 0.485 to 0.490, close to the ideal half-depth value $1/2$ associated with the energy-conserving gravity-current solution of~\citet{benjamin1968}.
For the dense front, comparison with values read from Fig.~5 of~\citet{birman2005} gives differences of approximately 0--3\% (panel b).
The reference calculation uses a longer tank, $32H$, and a different interface treatment, so the comparison is a check of front propagation rather than a matched discretisation study.
An additional no-slip calculation at $\gamma=0.4$ produces slower fronts, demonstrating sensitivity to the wall treatment.
Separating this effect from discretisation and domain-length effects requires grid- and tank-length sensitivity studies, which are outside the scope of this manuscript, before the remaining discrepancies can be attributed to a specific numerical mechanism.

\subsection{Wave propagation over a submerged bar}
\label{sec:bar}
Non-breaking propagation over a submerged bar assesses the ability of the wave model to reproduce bound-harmonic generation during shoaling and harmonic release over decreasing bed elevation.
The geometry follows the laboratory flume of~\citet{beji1993}, with $x$ measured from the wavemaker.
The still-water depth is 0.40 m, and the bar has a 1:20 upslope starting at $x=6$ m, a 2 m crest at 0.10 m depth between $x=12$ and 14 m, and a 1:10 downslope ending at $x=17$ m.
The incident wave has height $H=0.02$ m and period $T=2.02$ s, so that the offshore wavelength is $L=3.737$ m and $kh=0.67$.
The shoaling wave generates bound second and third harmonics over the crest, which become free waves on the downslope, so the surface elevation downstream of the bar contains several components of comparable amplitude~\citep{beji1993,beji1994}.

The inlet uses a stream-function wave with 16 terms and a ramp of two periods.
The generation zone is one wavelength long and the absorption zone two wavelengths long, with a relaxation rate of $20\,\mathrm{s^{-1}}$.
The domain is 33.47 m long and 0.5 m high, with 0.1 m of air above the still-water level, and is discretised by $3347\times250\times3$ cells with $\Delta x_1=10$ mm and $\Delta x_2=2$ mm, giving 10 cells over the incident wave height.
Periodic boundaries across the three spanwise cells give a quasi-two-dimensional calculation, for which no SGS closure is applied.
The bar is an immersed body described by its exact signed distance on the computational grid, and log-law wall models act on both the flat bed and the bar; the top boundary is free-slip.
The time step is adaptive with a Courant-number target of 0.2 (median $\Delta t=0.52$ ms), and the run lasts 50 s.
Harmonic amplitudes are obtained by a least-squares fit of the first three harmonics of the wave period to the last five periods of the surface-elevation gauges.

The experimental reference comprises regular-wave records from the Delft Hydraulics experiments, compiled by~\citet{dingemans1997}, at 11 gauges between $x=2$ and 21 m.
These records are distributed with the bar validation case in the wave-generation toolbox of~\citet{jacobsen2012}; they are distinct from the regular-wave signals presented by~\citet{beji1993} without calibrated amplitude or time axes.
Applying the same harmonic fit to the last four periods of each measured record provides the reference amplitudes and phases.
The simulation adopts the measured period of 2.02 s to ensure comparison at identical frequencies.
\begin{figure}[htbp]
  \centering
  \includegraphics[width=0.85\linewidth]{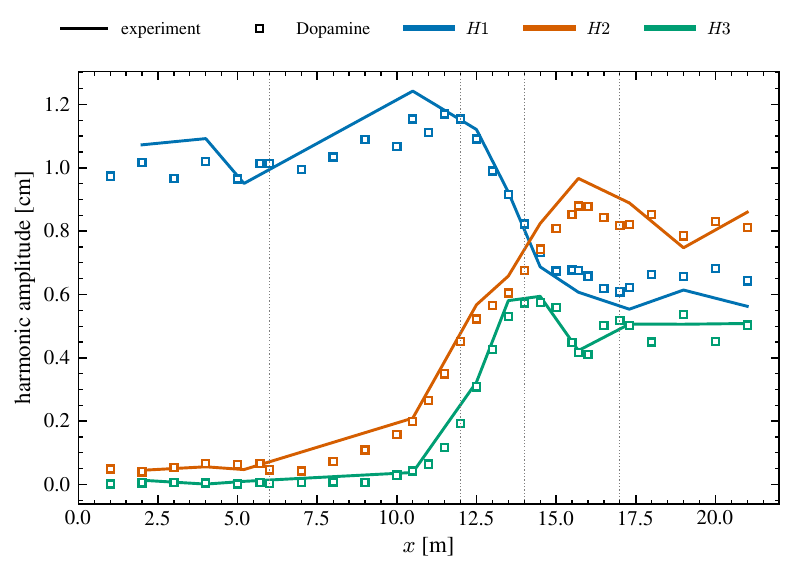}
  \caption{Amplitudes of the first three harmonics (H1--H3) of the surface elevation along the flume for $H=0.02$ m and $T=2.02$ s.
  Lines show the measurements and squares show Dopamine.
  Dotted vertical lines mark the corners of the bar.}
  \label{fig:bar-harmonics}
\end{figure}

The harmonic amplitudes in Fig.~\ref{fig:bar-harmonics} show the growth and redistribution of wave components across the bar.
At $x=2$ m, the simulated first-harmonic amplitude is 1.017 cm and the measured amplitude is 1.072 cm.
The first harmonic grows to 1.154 cm (measured 1.242 cm) on the upslope at $x=10.5$ m and decreases on the downslope as energy moves to the higher harmonics.
The second harmonic grows from 0.05 cm before the bar to a maximum of 0.879 cm at $x=15.7$ m, compared with a measured maximum of 0.966 cm at the same gauge.
The third harmonic reaches 0.574 cm, compared with 0.594 cm measured.
Across the 11 gauges, the RMS differences are 0.058, 0.051 and 0.020 cm for the first, second and third harmonics, equal to 6.5, 7.9 and 5.0\% of the corresponding RMS measured amplitudes, and the largest differences are 0.088, 0.087 and 0.050 cm.
The simulated first harmonic exceeds the measurement by 0.04--0.08 cm at $x\ge14.5$ m, whereas the second harmonic is lower than the measurement at 8 of the 11 gauges.

\begin{figure}[htbp]
  \centering
  \includegraphics[width=\linewidth]{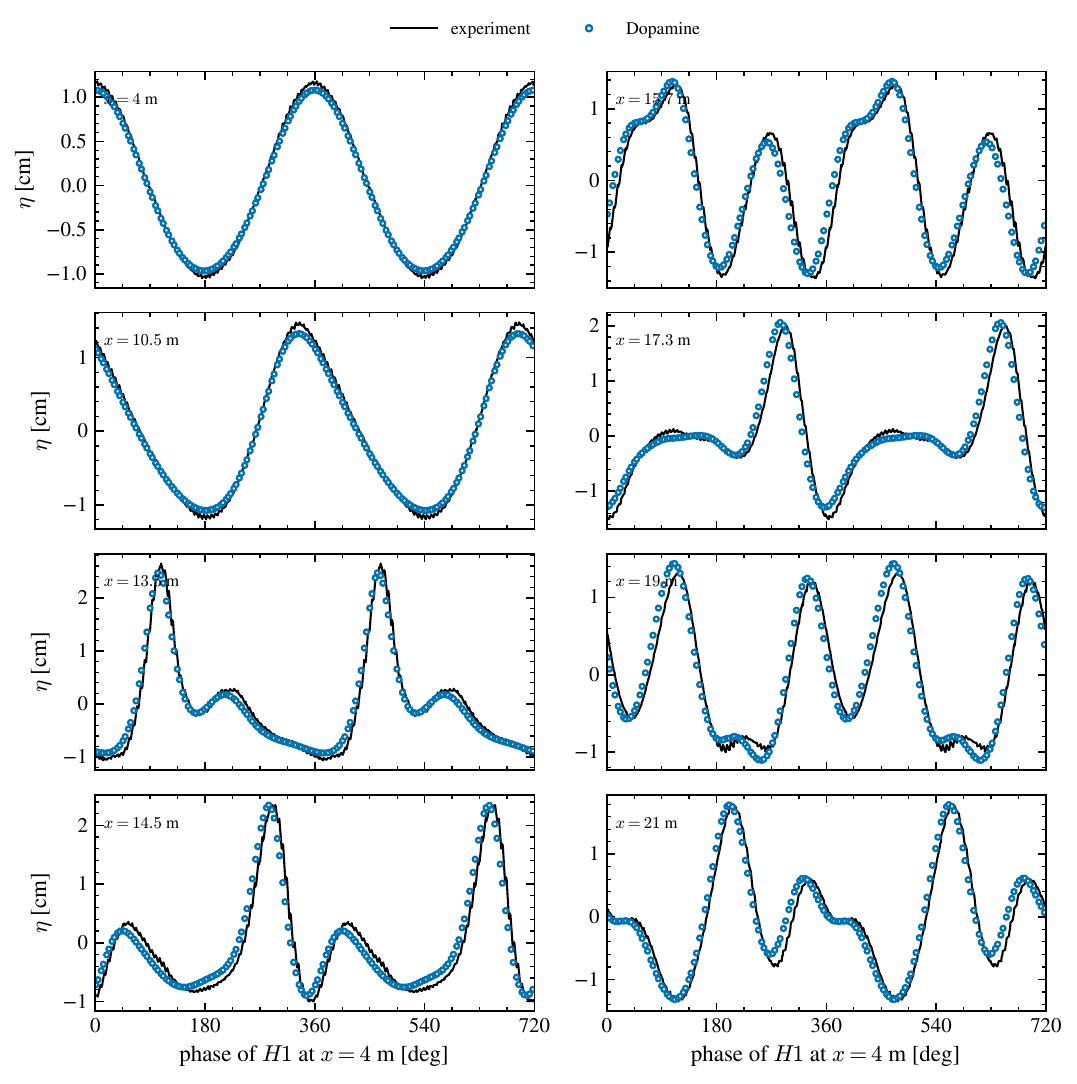}
  \caption{Surface elevation $\eta$ folded onto the wave period at eight gauges for $H=0.02$ m and $T=2.02$ s.
  The abscissa is the phase of the first harmonic at $x=4$ m, applied separately to the measured and simulated records, and the curves span four periods in the measurements and the last four periods in the simulation.
  Lines show the measurements and circles show Dopamine.}
  \label{fig:bar-series}
\end{figure}

The phase-folded surface-elevation records in Fig.~\ref{fig:bar-series} show how this harmonic evolution changes the wave shape.
The profile is nearly sinusoidal at $x=4$ and 10.5 m and develops a sharp crest at $x=13.5$ m.
Downstream of the crest, the released harmonics travel at their own phase speeds and the profile becomes multi-peaked for $x\ge15.7$ m.
The simulated records reproduce this evolution of shape at every gauge, with small differences in the trough levels and the secondary maxima.
The phase of the first harmonic relative to that at $x=2$ m differs from the measurement by at most 6.4$^\circ$ for $x\ge10.5$ m.

The reflection coefficient from the gauges at $x=2$ and 3 m is 0.03, the change in first-harmonic amplitude between the first and second half of the analysis window is at most 0.07 cm, and the liquid-volume residual of Eq.~\eqref{eq:ledger} does not exceed $6\times10^{-12}$ m$^3$.
Together, the harmonic and waveform comparisons support the reproduction of bound-harmonic generation and release, with harmonic-amplitude differences below approximately 0.09 cm.
Because the calculation uses a single grid and the measurements have no uncertainty estimate, these differences cannot be separated into discretisation, modelling and experimental contributions, nor can their convergence be established.

\subsection{Turbulent wave breaking validation}
The final case assesses the ability of Dopamine to model wave transformation, breaking and the associated mean flow for the spilling breaker of~\citet{ting1994}.
Waves of period 2 s and height 0.125 m propagate from a still-water depth of 0.40 m onto a 1:35 beach.
The wave tank has dimensions $20 \times 0.8 \times 0.19$~m in the streamwise, vertical, and spanwise directions, respectively.
The domain is discretised using $2000\times161\times20$ cells, and the Vreman sub-grid-scale model is used~\citep{vreman2004}.
The sloping bed is introduced using GenSDF~\citep{patilgarciasanchez2025gensdf} with a log-law wall model on the no-slip boundaries.
A stream-function inlet supplies the incident waves using a relaxation zone to avoid spurious reflections, which may contaminate the incoming wave signal.
The governing equations are integrated for a total of 60 wave periods, with statistics accumulated over 40 wave periods following an initial spin-up of 20 wave periods.

\begin{figure}[htbp]
  \centering
  \includegraphics[width=\linewidth,height=\linewidth,keepaspectratio]{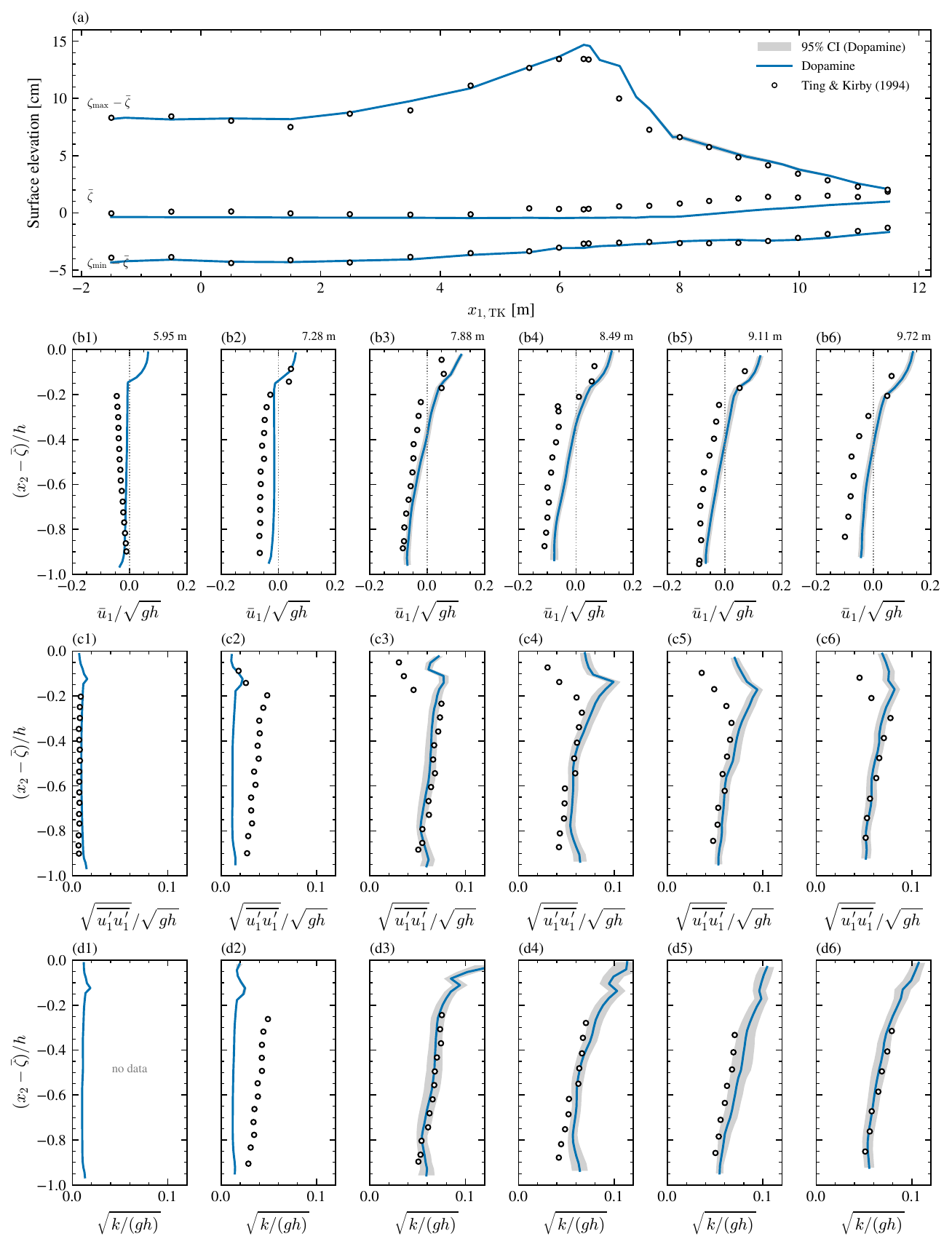}
  \caption{Numerical validation of wave transformation and turbulence under breaking waves: (a) surface elevations; (b) mean velocity; (c) streamwise RMS fluctuations; (d) the two-component kinetic-energy proxy at six stations.
  Curves and grey nominal 95\% sampling bands use 40 periods after $t=40$ s.
  Circles show \citet{ting1994} measurements.}
  \label{fig:tk}
\end{figure}

Fig.~\ref{fig:tk} compares the surface elevations and velocity statistics with experimental data from~\citet{ting1994}.
The surface elevations compare well with the experimental data and correctly capture the wave transformation as it reaches the nearshore, including wave dissipation due to wave breaking and subsequent roller development.
The velocity profiles also agree reasonably well with the experimental data, capturing the overall trend, with some discrepancies near the interface, as expected due to the limited vertical resolution.
The turbulent component also compares well with the experimental data, which supports the use of the framework for breaking-wave turbulence at this resolution.

\begin{figure}[htbp]
  \centering
  \includegraphics[width=\linewidth]{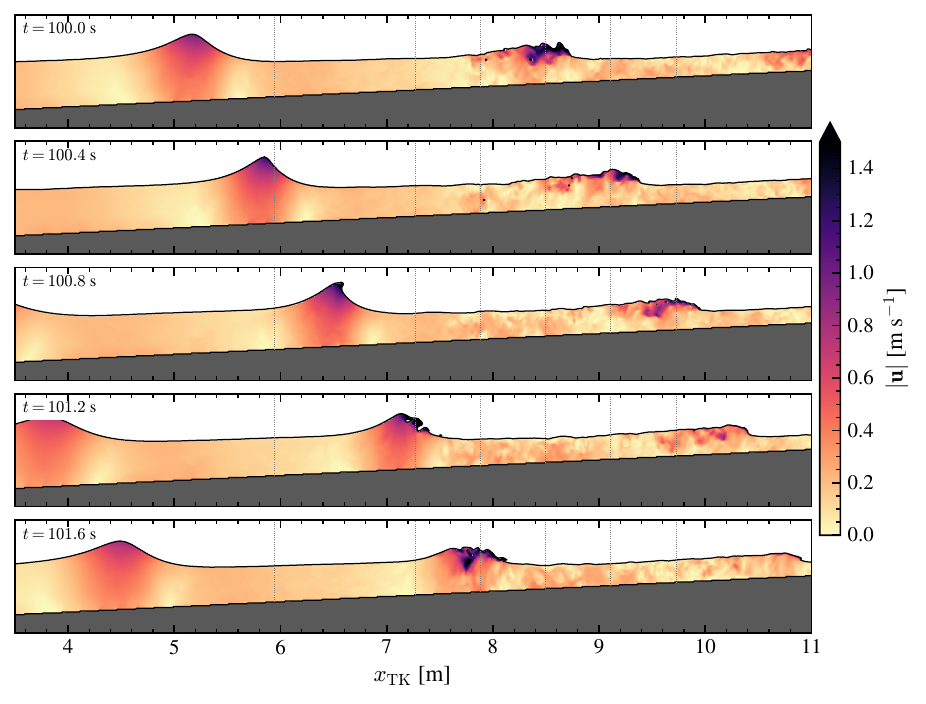}
  \caption{Streamwise mid-plane slices of the speed $|\mathbf{u}|$ in the liquid for the spilling breaker at five instants separated by 0.4~s, with the $C=0.5$ contour in black and the sloping bed in grey.
  Dotted vertical lines mark the six measurement stations of Fig.~\ref{fig:tk}.
  The time $t$ is measured from the start of the simulation, and the slices are taken at the spanwise mid-plane.}
  \label{fig:tk-slices}
\end{figure}

Fig.~\ref{fig:tk-slices} shows streamwise mid-plane slices of the speed in the liquid through one wave period.
The crest steepens as it propagates from $x_{\mathrm{TK}}\approx5.2$ to 6.5 m ($t=100.0$--$100.8$ s), and the front face collapses between $x_{\mathrm{TK}}\approx7$ and 8 m ($t=101.2$--$101.6$ s), where the largest speeds ($|\mathbf{u}|>1.4\,\mathrm{m\,s^{-1}}$) occur at the toe of the breaking crest.
Downstream, the bore generated by the preceding breaking event advances over the measurement stations with speeds of 0.3--0.6~$\mathrm{m\,s^{-1}}$ and fine-scale variations in speed, consistent with the turbulent fluctuations in Fig.~\ref{fig:tk}(c,d).

\subsection{Parallel performance}
\label{sec:scaling}
The parallel performance of the CPU solver is measured for the two-fluid configuration, using the breaker flume of the previous subsection with a flat wall-modelled bed and no immersed body.
The case uses a stream-function wave ($H=0.125$ m, $T=2$ s, $h=0.4$ m), the Vreman model, mass--momentum-consistent advection and the variable-coefficient pressure solver with a multigrid preconditioner.
The choice of the sub-grid model does not affect the overall performance of the solver, thus only the Vreman sub-grid model was tested as the sub-grid contributes only a marginally small fraction of the computational cost per time step.
The grid spacing is $\Delta x_1=\Delta x_3=10$ mm and $\Delta x_2=5$ mm, and the 160 wall-normal cells are never decomposed.
Each run advances 30 time steps from still water without file output, and the time per step is the mean wall time over steps 6--30.
The runs use 16 to 512 MPI ranks, one rank per core, on exclusive nodes of 64 cores each, with one run per configuration.
Strong scaling uses a fixed $2049\times160\times65$ grid ($2.1\times10^{7}$ cells), and weak scaling keeps $64\times160\times64$ cells (approximately $6.6\times10^{5}$) per rank.
The pencil decomposition is prescribed so that the local $x_1$--$x_3$ blocks are nearly square (Table~\ref{tab:scaling}), and the mean number of pressure iterations is approximately 4.5 per step at every rank count, so the solver work per step does not change with the number of ranks.
Because a single node is filled between 16 and 64 ranks, efficiencies are quoted relative to 64 ranks unless stated otherwise.
The default multigrid cycle on CPUs is the W-cycle, and the V-cycle is available as an option.

\begin{table}[htbp]
  \centering
  \caption{Strong scaling of the two-fluid solver for a $2049\times160\times65$ grid.
  The efficiency is $64\,t_{64}/(P\,t_P)$, where $t_P$ is the time per step on $P$ ranks, and the pencil grid is $p_{\mathrm{row}}\times p_{\mathrm{col}}$ in $x_1$ and $x_3$.}
  \label{tab:scaling}
  \footnotesize
  \setlength{\tabcolsep}{4pt}
  \begin{tabular}{@{}rrrrrr@{}}
    \toprule
    Ranks & Pencil grid & Cells/rank & $t_P$ (W) [s] & $t_P$ (V) [s] & Efficiency (W / V)\\
    \midrule
    16  & $16\times1$  & $1.33\times10^{6}$ & 2.63 & 2.61 & ---\\
    32  & $32\times1$  & $6.7\times10^{5}$  & 1.79 & 1.77 & ---\\
    64  & $32\times2$  & $3.3\times10^{5}$  & 1.59 & 1.57 & 100\% / 100\%\\
    128 & $64\times2$  & $1.7\times10^{5}$  & 0.90 & 0.85 & 88\% / 93\%\\
    256 & $64\times4$  & $8.3\times10^{4}$  & 0.63 & 0.46 & 63\% / 85\%\\
    512 & $128\times4$ & $4.2\times10^{4}$  & 0.37 & 0.25 & 54\% / 77\%\\
    \bottomrule
  \end{tabular}
\end{table}

\begin{figure}[htbp]
  \centering
  \includegraphics[width=\linewidth]{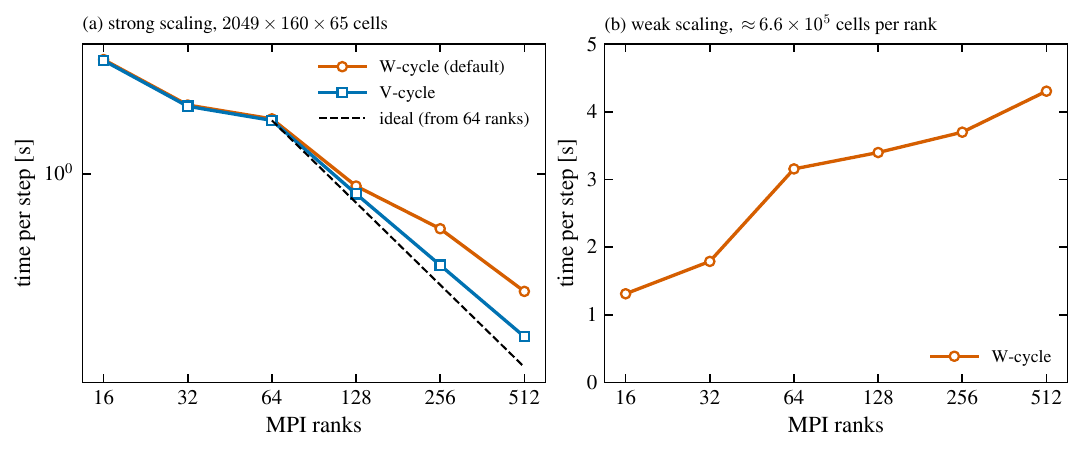}
  \caption{Scaling of the two-fluid CPU solver on 16 to 512 MPI ranks: (a) strong scaling for a fixed $2049\times160\times65$ grid with the W-cycle and V-cycle multigrid preconditioner; (b) weak scaling at approximately $6.6\times10^{5}$ cells per rank with the W-cycle.
  The dashed line in (a) shows ideal scaling from 64 ranks, and each point is a single run.}
  \label{fig:scaling}
\end{figure}

Fig.~\ref{fig:scaling} and Table~\ref{tab:scaling} show that the time per step falls from 2.63 s on 16 ranks to 0.37 s on 512 ranks with the W-cycle and to 0.25 s with the V-cycle, speed-ups of 7.1 and 10.3 against an ideal factor of 32.
Relative to 64 ranks, the strong-scaling efficiency at 512 ranks is 54\% for the W-cycle and 77\% for the V-cycle, with only $4.2\times10^{4}$ cells per rank.
Most of the loss occurs within a single node: with the same $6.6\times10^{5}$ cells per rank, the weak-scaling time rises from 1.31 s on 16 ranks to 3.16 s on 64 ranks, and the weak-scaling efficiency relative to 64 ranks is then 93, 85 and 73\% at 128, 256 and 512 ranks.
The intra-node loss is consistent with saturation of the memory bandwidth by the stencil kernels, although no hardware counters were collected to confirm this.
Across nodes, the multigrid preconditioner is the main limitation of the W-cycle, accounting for approximately 32\% of the profiled time at 512 ranks in the strong-scaling case, and the V-cycle reduces the time per step by a factor of 1.46 at this rank count with a similar number of pressure iterations (5.5 against 4.6 at 512 ranks).
The Poisson transposes are the next most costly stage at scale, and in weak scaling their time increases from 14 s on 64 ranks to 31 s on 512 ranks over the 30 steps.

The single-GPGPU build offloads the same time step to one NVIDIA A100 (80 GB) with OpenACC and was timed on the same flat-bed configuration with $1024\times160\times N_3$ cells, $N_3=16$, 32 and 64.
On $1024\times160\times64$ cells ($1.07\times10^{7}$), a single GPGPU advances a step in approximately 0.60 s, compared with 1.31 s on 16 CPU ranks, a speed-up of 2.2, and it needs a single process and device rather than a CPU node.
The pressure-solution stages (FFT, tridiagonal and PCG) and the volume-fraction advection are accelerated by a factor of 4--5, whereas the interface-force stage is accelerated by only 1.5 and has become the largest contribution (approximately 48\%) to the GPGPU time per step.
The time per step grows from 0.25 s to 0.37 s and 0.60 s for $2.6\times10^{6}$, $5.2\times10^{6}$ and $1.05\times10^{7}$ cells, so the cost remains proportionate to the problem size.
These GPGPU timings are single runs that exclude the initial device set-up, and the GPGPU build is limited by the memory of one device and has not been tested on multiple devices.

\section{Conclusions}
\label{sec:concluding-remarks}
We have presented Dopamine, a Cartesian staggered-grid large-eddy-simulation framework for free-surface flows that combines geometric volume-of-fluid transport, consistent mass--momentum advection, a variable-density pressure projection, ghost-cell immersed boundaries and one-way-coupled Lagrangian particle tracking.
Verification against analytical solutions and validation against laboratory and numerical benchmarks show that the solver preserves hydrostatic and capillary equilibrium across large density contrasts, transports the interface in a bounded and volume-conserving manner, and reproduces density-driven fronts, particle response and the transition of buoyant particles through the free surface.
For wave propagation over a submerged bar and wave breaking on a plane beach, the simulated harmonics, surface elevations, mean flow and turbulence are in good overall agreement with the measurements, with the largest differences near the interface.
Together with a favourable parallel efficiency on several hundred CPU cores, these results indicate that Dopamine is a suitable low-numerical-dissipation tool for investigating small-scale turbulent physics in coastal free-surface flows.

Future development of Dopamine will concentrate on three areas.
First, the performance of the solver will be extended beyond the present single-device and CPU results: the interface-force stage, which limits the gain of the current single-GPGPU build, will be fully offloaded, the build will be developed into a multi-GPGPU implementation so that larger domains can be simulated, and the communication-bound multigrid and transpose stages of the pressure solution will be optimised to improve efficiency at high rank counts.
Second, the geometric accuracy at immersed boundaries will be improved, since the pressure and volume-fraction treatments currently use staircase faces, so that complex coastal bathymetry and structures can be represented more accurately.
Third, the wave generation will be extended to irregular sea states that retain nonlinear bound harmonics, which the present record-driven formulation does not reconstruct.

\section*{CRediT authorship contribution statement}
Akshay Patil: Conceptualization, Methodology, Software, Validation, Formal analysis, Investigation, Data curation, Writing - original draft, Writing - review \& editing, Visualization.
\section*{Declaration of competing interest}
The author declares that they have no known competing financial interests or personal relationships that could have appeared to influence the work reported in this paper.
\section*{Open Research Section}
The source code, example case configurations, documentation and selected validation summaries are available in the public Dopamine repository at \url{https://github.com/AkshayPatil1994/dopamine-fdm}.
Additional documentation is provided in the repository wiki at \url{https://github.com/AkshayPatil1994/dopamine-fdm/wiki}.
\section*{Use of Generative-AI and AI-assisted technologies}
During the preparation of this work, the author used Grammarly to improve the text's clarity and readability. Additionally, Anthropic's Sonnet-5.2 and 5.5 large language models (LLM) were used to assist with code development and debugging. The use of LLMs was restricted to code audit for CPU performance improvement, subroutine implementations after an initial draft, GPGPU porting efforts, and documentation. The author reviewed both the text and the code and takes full responsibility for the publication's content. A transparent record of the AI-assisted technologies used in this work is maintained in the source code's git repository through the commit history and the issues raised during development.
\section*{Acknowledgements}
This work was supported by the Climate Action Fund of the Delft University of Technology.
The author acknowledges the use of computational resources of the DelftBlue supercomputer, provided by Delft High Performance Computing Centre (https://www.tudelft.nl/dhpc).
The author thanks Clara Garc\'{i}a-S\'{a}nchez and Hugo Ledoux for their support and guidance throughout the project.
The author also thanks Chit Yan Toe for testing the Dopamine particle module and providing feedback on the usability of the code and its documentation.

\appendix

\bibliographystyle{elsarticle-harv}
\bibliography{references}
\end{document}